\documentclass[11pt]{article}
\usepackage{moriond,epsfig}
\usepackage{amssymb}

\def\Journal#1#2#3#4{{#1} {\bf #2}, #3 (#4)}

\def\PRD{{\em Phys. Rev.} D}

\def\be{\begin{equation}}
\def\ee{\end{equation}}
\def\bea{\begin{eqnarray}}
\def\eea{\end{eqnarray}}

\begin{document}
\vspace*{4cm}
\title{VECTOR MESON PRODUCTION AT HERA}

\author{ K.-C. VOSS }

\address{Physikalisches Institut der Universit\"at Bonn,\\
Nussallee 12, 53115 Bonn, Germany}

\maketitle\abstracts{
At the HERA collider the experiments H1 and ZEUS have studied the
exclusive production of vector mesons over a wide kinematical range. The
recent measurements and their discussion within the framework of color
dipole models and pQCD are reported.  
}

\section{Introduction}
Exclusive photo- and electroproduction of light ($\rho,\omega,\phi$)
and heavy ($J/\psi,\psi^{\prime},\Upsilon$) vector mesons (VM)
have been  subject  of intensive studies at HERA. The accelerator and its general purpose detectors H1 and ZEUS provide a unique opportunity to measure the exclusive diffractive production of vector mesons with different masses $M_{VM}$ in photo- and electroproduction.

\subsection{Diffractive vector meson production}
\begin{center}
\begin{figure}[h]
\begin{center}
\epsfig{file=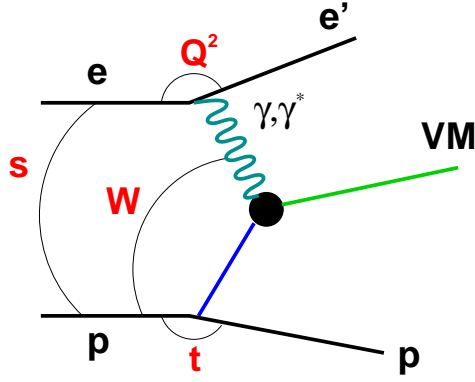,height=5cm}
\end{center}
\caption{Diagram of exclusive vector meson (VM) production at HERA.}
\label{fig:diagram}
\end{figure}  
\end{center}

The process $ep \rightarrow e (VM) p$, drawn in fig.~\ref{fig:diagram},  can be described as  a two step process. The incoming electron emits a photon. This photon fluctuates into a $q\bar{q}$ state which scattters with the proton by exchanging nothing but momentum.
The kinematical variables which are used to characterize the process are: the 4-momentum transfer squared $Q^{2}$ at the electron vertex, the center of mass energy of the $\gamma$-proton system $W$ and the 4-momentum transfer squared $t$ at the proton vertex. The dependences of the cross sections on these variables are presented in the ranges:  $2 < Q^{2} < 100$ $GeV^2$, $30 < W < 260$ $GeV$ and $|t|<1$ $GeV^2$.

In the  absence of a hard scale the  colorless exchange between the vector meson and the proton can be
modeled by a soft pomeron trajectory using the Regge approach. In this
approach the cross section is predicted to rise slowly with
$W$. In the presence of a hard scale, the vector meson
production can be calculated using pertubative QCD (pQCD). In this case the 
colorless exchange is modeled in leading order by a pair of
gluons, the cross section is proportional to the square of
the gluon density. This  predicts a steep rise  with increasing values of $W$.

\section{$W$ dependence}

For exclusive photoproduction ($Q^{2} \approx 0$ $GeV^2$) of vector
mesons the $W$ dependence is shown in fig.2a). 
The lines indicate the rising of the
cross sections assuming the form $W^{\delta}$. For the light vector mesons
 $\rho$, $\omega$ and $\phi$ the slope is $\delta \approx
0.22$. This value is very similar to the total photoproduction cross
section and is predictied by the Regge approach.  For the heavier
vector mesons $J/\psi$, $\psi(2s)$ and $\Upsilon$ the observed slope
is higher ($\delta \gtrsim 0.8$) .

\begin{figure}[h]
  \begin{minipage}{\textwidth}
    \begin{tabular}{cc}
      \begin{minipage}{0.5\textwidth}
a)\\
	\epsfig{file=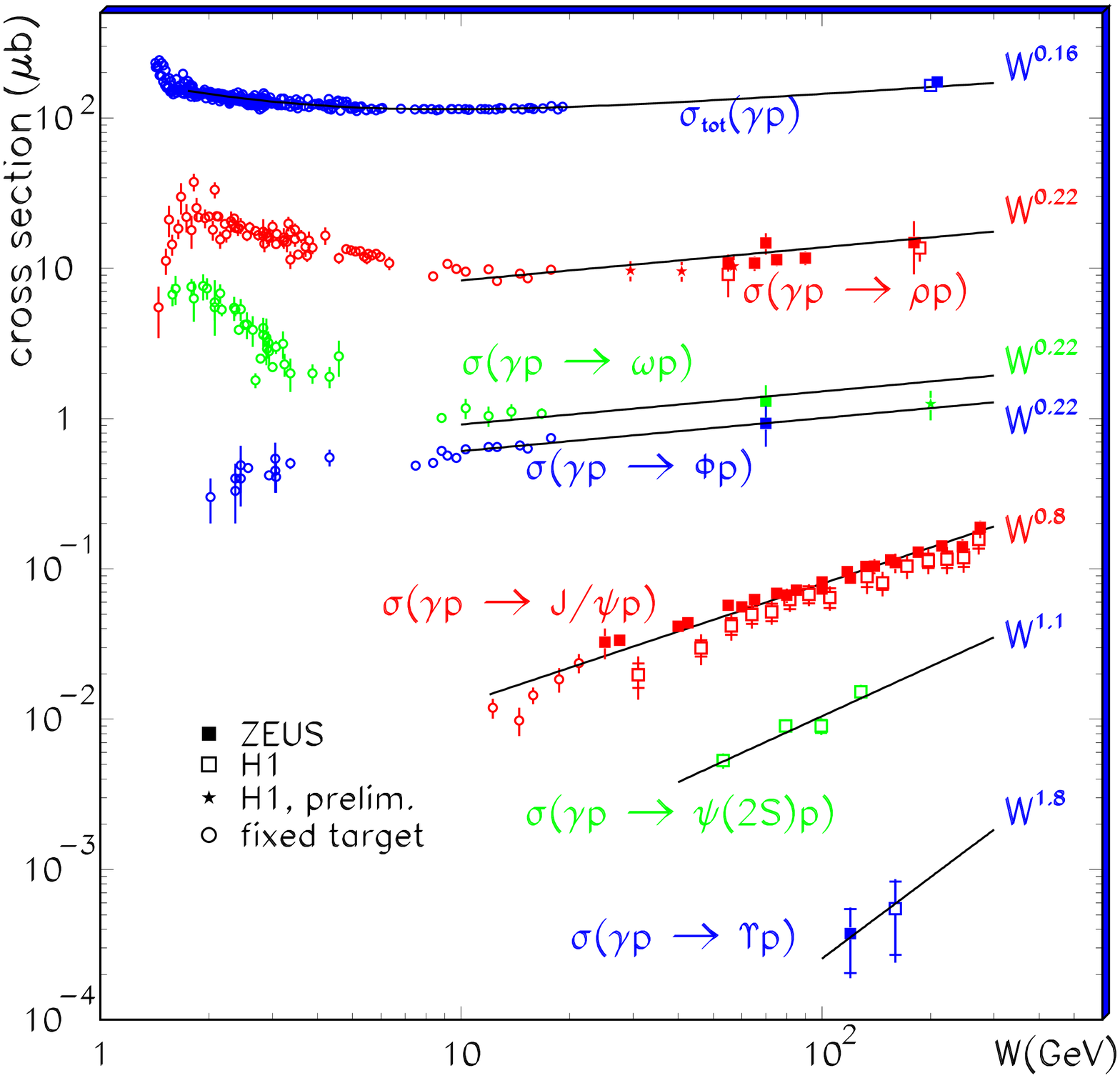,height=7.5cm}
      \end{minipage}
      &

      \begin{minipage}{0.5\textwidth}
b)\\
	\epsfig{file=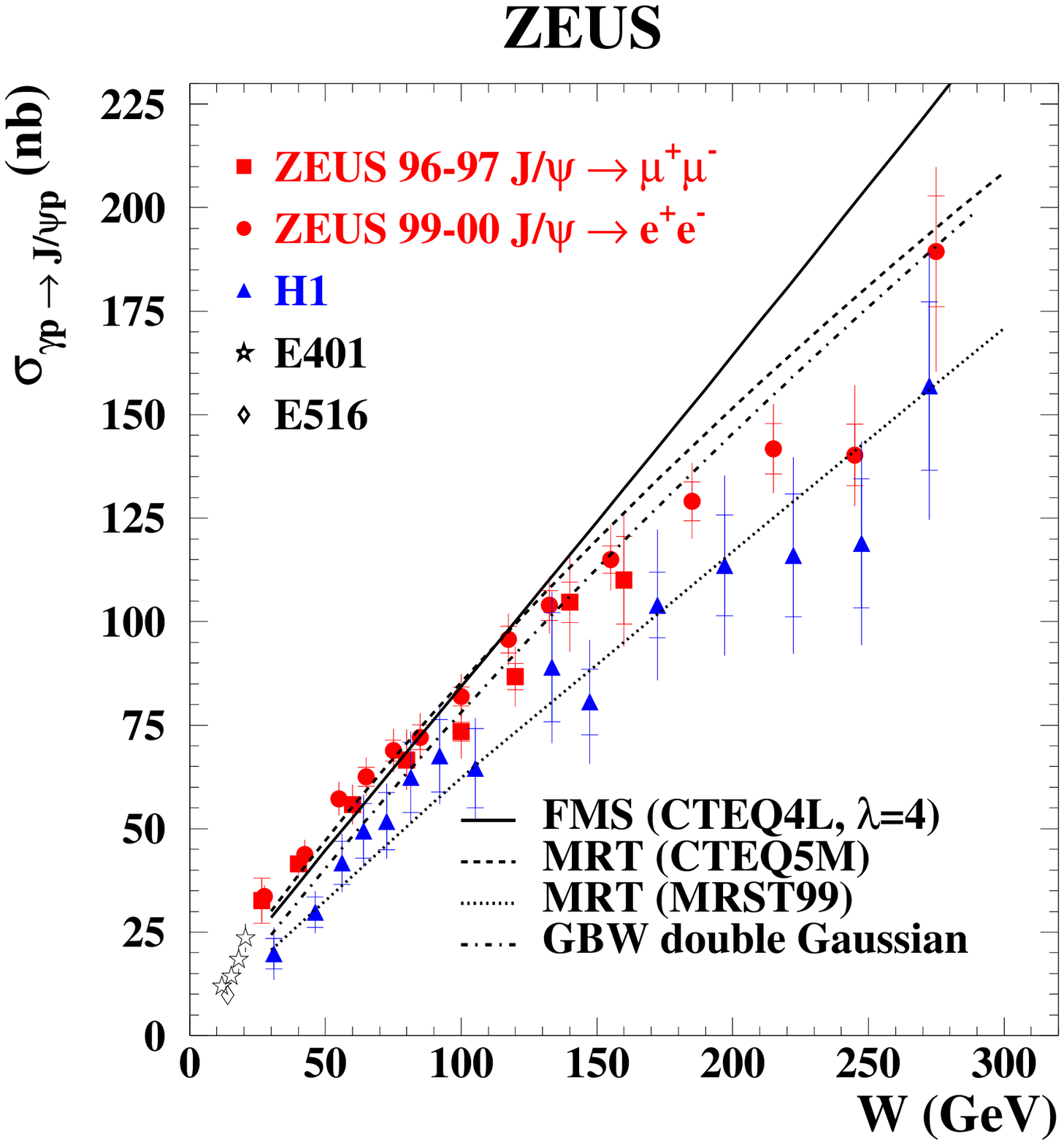,height=8cm}
      \end{minipage}
    \end{tabular}
  \end{minipage}
  
  \caption{a) Compilation of $W$ dependences for exclusive
  photoproduction of light and heavy vector mesons. The lines show the
  behaviour of the cross section assuming the form $W^{\delta}$. The values
  for $\delta$ are given at the right edge. 
b) $W$ dependence of the $J/\psi$  photoproduction cross
  section in comparison with theoretical models based on  pQCD.}
  \label{fig:php}
\end{figure}

For the heavy vector mesons the masses of the charm and the bottom quarks  provide a hard scale  which allows the use of pQCD to calculate the cross sections.
Such models are able to describe the $J/\psi$~\cite{phpjpsiZEUS,phpjpsiH1} cross section as it is shown in fig.2b). In particular the steep rise as a function of the energy $W$ is well described by pQCD~\cite{theory}.

\section{$Q^2$ and $|t|$ dependence}
The $|t|$ dependence of the cross section for exclusive vector meson
production is well described by the form $\frac{d\sigma}{dt} \propto e^{-b|t|}$
for small values of $t$ ($|t| <1$ $GeV^2$). Fig.3a+b) show the
dependence of the slope $b$ as a function of $Q^2$ for
$\rho$~\cite{rho-elastic} and $J/\Psi$~\cite{disjpsiZEUS} in photo-
and electroproduction. The slope of the $\rho$ cross section decreases
with increasing $Q^2$. This indicates that the size of the interacting region is
changing with $Q^2$.  
On the right hand side one can see that, in contrast to the $\rho$, the $J/\psi$ has is no change in the slope with $Q^2$. The production mechanism for $J/\psi$  at
the photoproduction limit is already the same
as in the higher $Q^2>0$ range. This
is interpreted as due to the fact that the $J/\psi$ mass alrady provides
a hard scale at $Q^2=0$, in contrast to  
exclusive $\rho$ electroproduction.
\begin{figure}[t]
  \begin{minipage}{\textwidth}
    \begin{tabular}{cc}
      \begin{minipage}{0.5\textwidth}
a)\\
	\epsfig{file=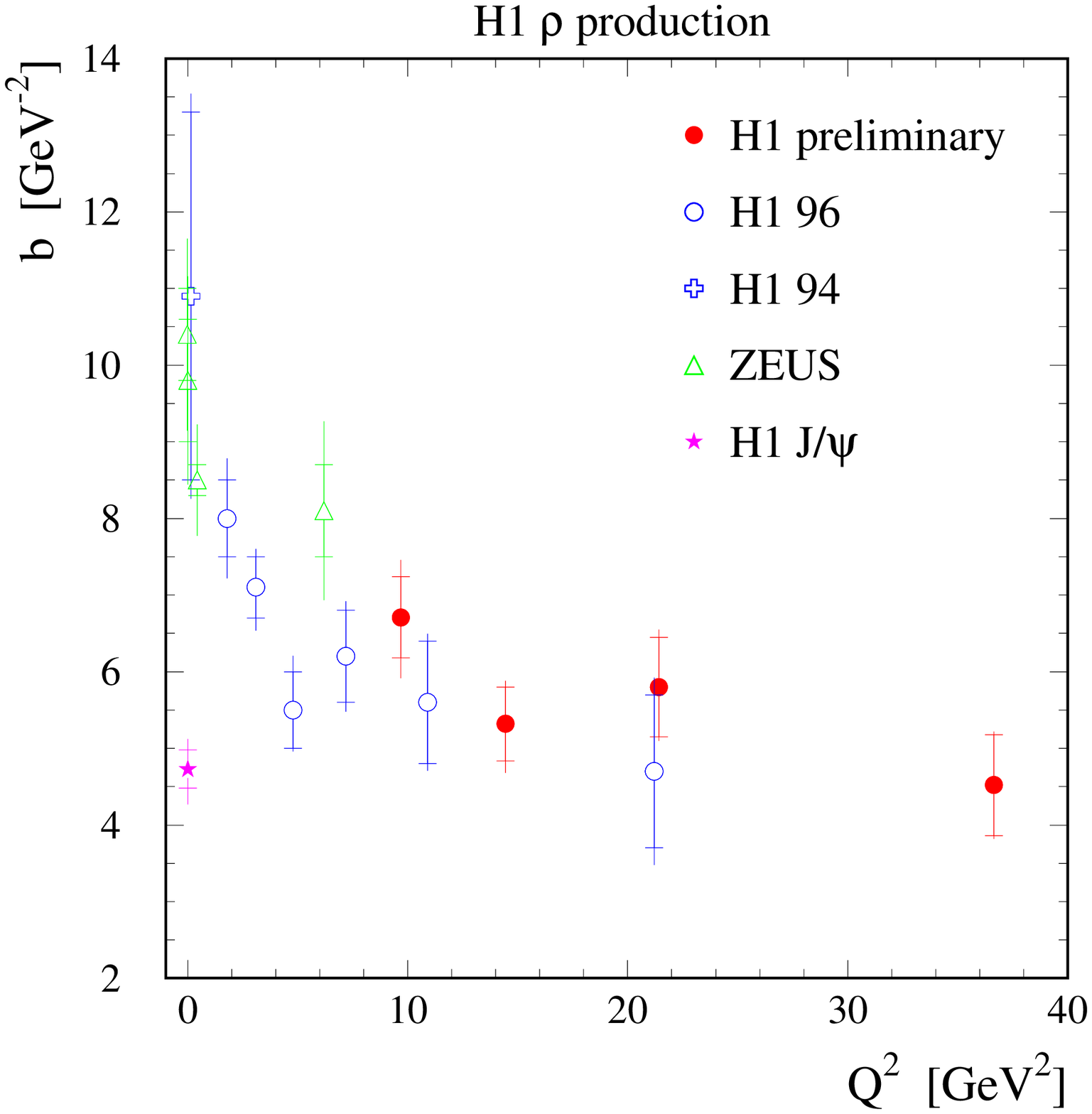,height=6cm}
      \end{minipage}
      &

      \begin{minipage}{0.5\textwidth}
b)\\
	\epsfig{file=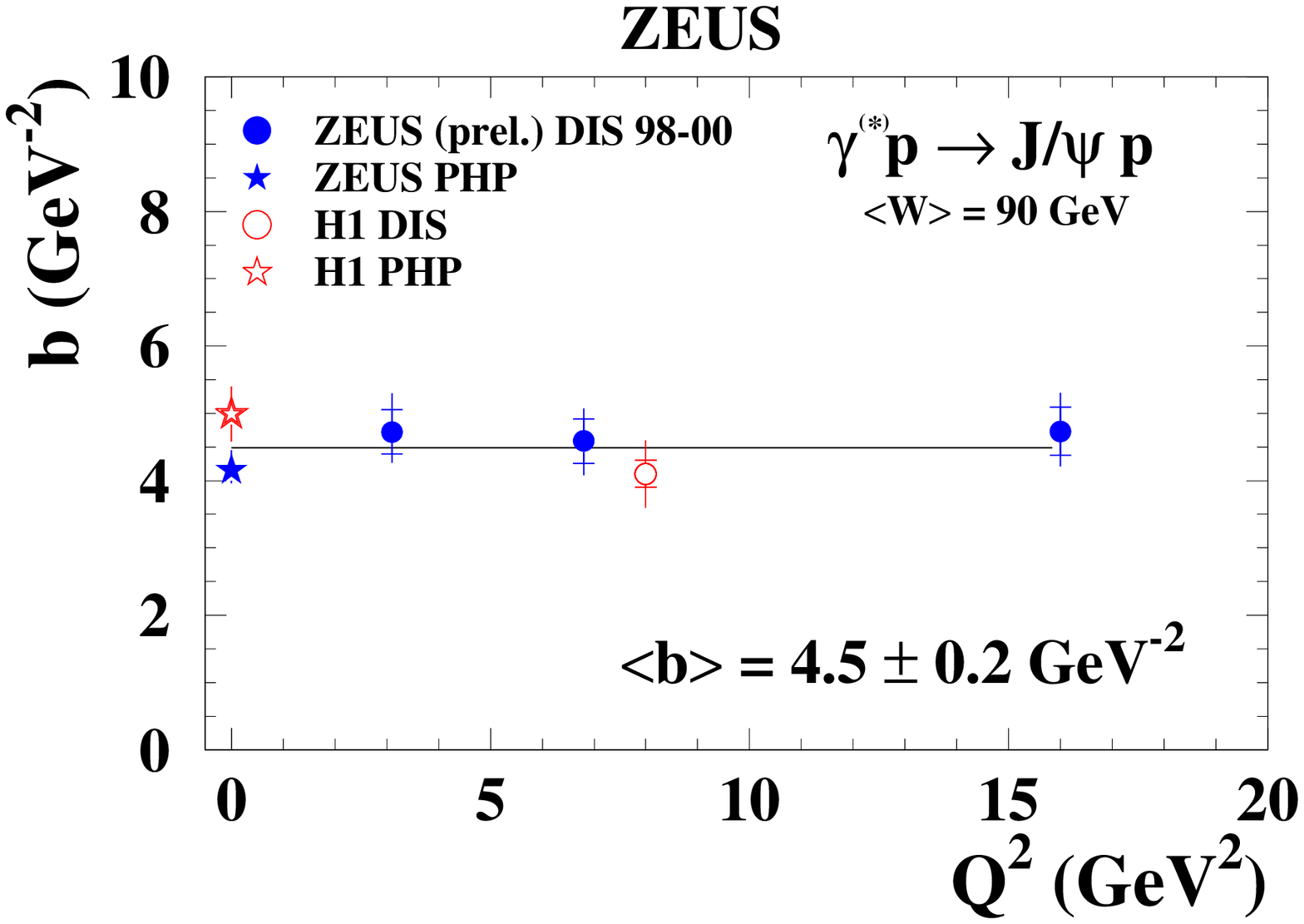,height=6cm}
      \end{minipage}
    \end{tabular}
  \end{minipage}
 
  \caption{a) Fitted values of the $b$-slope as a function of $Q^2$ in
  $\rho$ photo- and electroproduction.
  b) Fitted values of the $b$-slope as a function of $Q^2$ in
  $J/\psi$ photo- and electroproduction.}
  \label{fig:php}
\end{figure}  

In the diffractive picture $b$ is related to the radii of the
colliding objects i.e. of the proton and the VM: $b\propto r^2_{VM}
+R^2_{\rm proton}$. The values of $b \approx 4.5$ $GeV^{-2}$ measured
in the hard regime implies a combined radius of the order of the size
of the proton. This observation suggests that the transverse size of
the $q\bar{q}$ fluctuation producing the $J/\psi$ in photoproduction
and the $\rho$ at high $Q^2$ is smaller than that of the
proton.

\section{Decay angular distributions}

The production  and decay of  a  VM into a pair of oppositely 
charged particles  can be described
in terms of three angles: $\Phi_h$,  the angle between the VM production plane
and the lepton scattering plane; $\theta_h$ and $\phi_h$,  the polar
and azimuthal  angles of the positively charged decay lepton in the
$s$-channel helicity  frame.
\begin{figure}[h]
  \begin{minipage}{\textwidth}
    \begin{tabular}{cc}
      \begin{minipage}{0.5\textwidth}
a)\\
	\epsfig{file=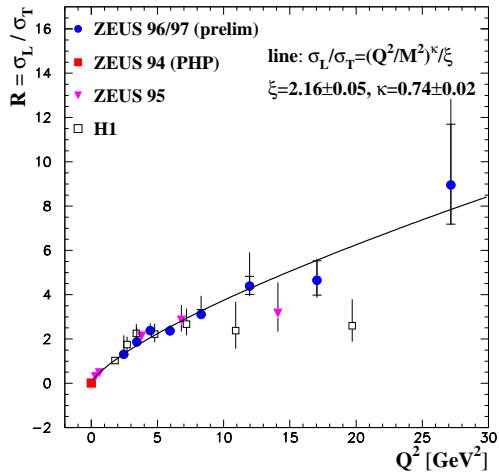,height=7cm}
      \end{minipage}
      &

      \begin{minipage}{0.5\textwidth}
b)\\
	\epsfig{file=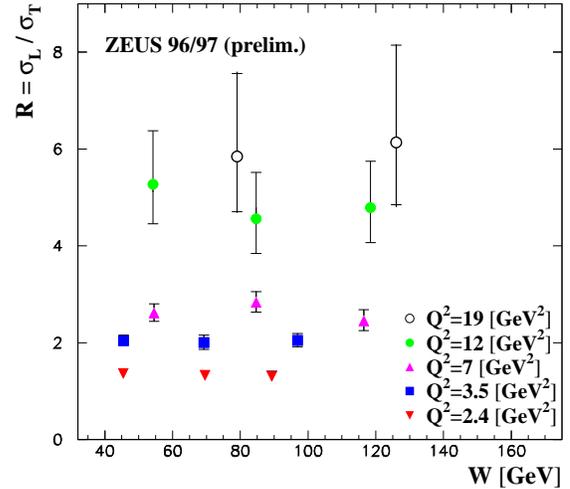,height=8cm}
      \end{minipage}
    \end{tabular}
  \end{minipage}
  
  \caption{a) $Q^2$ dependence of $R$ in  $\rho$ electroproduction.
b) $W$ dependence of $R$ in  $\rho$ electroproduction in
different bins of $Q^2$.}
  \label{fig:php}
\end{figure}  

Under the assumption of $s$-channel helicity conservation (SCHC), the
angular distribution for the decay of the VM depends only on two
angles, $\theta_h$ and $\psi_h = \phi_h - \Phi_h$.  From the
$\theta_h$ distribution the spin density matrix element $r^{04}_{00}$
can be extracted which is proportional to the the helicity amplitude
$T_{00}$.  $T_{00}$ corresponds to an amplitude, where a
longitudinally polarized photon yields a longitudinally polarized
VM. Also under the assumption of SCHC the ratio $R$ of cross sections
for longitudinally and transverse polarized photons can be calculated
using $r^{04}_{00}$; ($R = \frac{1}{\epsilon}
\frac{r^{04}_{00}}{1-r^{04}_{00}}$). This measurement for $\rho$
electroproduction is shown in fig.4). The left plot shows the rise of
$R$; it indicates that the longitudinal dipole configuration becomes
more dominant when increasing the value of $Q^2$. Within pQCD models a
small dipole is most likely to be produced if the virtual photon is
longitudinally polarized, which predicts that $|T_{00}|$ is the
dominant amplitude.

Fig.4b) shows the ratio $R$ in $\rho$ electroproduction for
different values of $Q^2$ regions as a function of $W$.
The data is consistent with no  $W$ dependence. 

\section*{Acknowledgments}

I would like to thank the organizers of the very interesting and diverting conference as well as my colleagues at the H1 and ZEUS collaboration for providing the figures for this talk.

\end{document}